\documentclass[aps,prl,reprint,twocolumn,groupedaddress,superscriptaddress,nofootinbib]{revtex4-1}

\usepackage{graphics}
\usepackage[mathcal]{euscript}
\usepackage[latin1]{inputenc}
\usepackage{latexsym}
\usepackage{amsmath}
\usepackage{hyperref}   
\usepackage{graphicx}   
\usepackage{verbatim}  
\usepackage{color}      
\usepackage{subfigure}  
\usepackage{hyperref}   
\usepackage{bm}
\usepackage{graphicx}
\usepackage{amssymb}
\usepackage{bbm}
\usepackage{float}
\usepackage{slashed}
\usepackage{amsfonts}
\usepackage{euscript}
\usepackage{mathrsfs} 
\usepackage{bbding}
\usepackage{pifont}
\usepackage{cancel}
\usepackage{amssymb,fge}
\usepackage{pifont}
\newcommand{\cmark}{\ding{51}}%
\newcommand{\xmark}{\ding{55}}%
\newcommand{\ii}{\mathrm{i}}

\usepackage[usenames,dvipsnames]{xcolor}
\hypersetup{colorlinks=true, citecolor=Magenta, linkcolor=Magenta,
urlcolor=Magenta}

\usepackage{tikz}
\usetikzlibrary{decorations.pathmorphing,decorations.markings}

\begin{document}

\title{The Unruh effect without thermality}


\author{Ra\'ul Carballo-Rubio}
\email{raul.carballorubio@sissa.it}
\affiliation{SISSA, International School for Advanced Studies, Via Bonomea 265, 34136 Trieste, Italy}
\affiliation{INFN Sezione di Trieste, Via Valerio 2, 34127 Trieste, Italy} 
\author{Luis J. Garay}
\email{luisj.garay@ucm.es}
\affiliation{Departamento de F\'{\i}sica Te\'orica, Universidad Complutense de Madrid, 28040 Madrid, Spain}
\affiliation{Instituto de Estructura de la Materia (IEM-CSIC), Serrano 121, 28006 Madrid, Spain}
\author{Eduardo Mart\'in-Mart\'inez}
\email{emartinmartinez@uwaterloo.ca}
\affiliation{Institute for Quantum Computing, University of Waterloo, Waterloo, ON, N2L 3G1, Canada}
\affiliation{Dept. Applied Math., University of Waterloo, Waterloo, ON, N2L 3G1, Canada}
\affiliation{Perimeter Institute for Theoretical Physics, Waterloo, ON, N2L 2Y5, Canada}
\author{Jos\'e de Ram\'on}
\email{jderamonrivera@uwaterloo.ca}
\affiliation{Institute for Quantum Computing, University of Waterloo, Waterloo, ON, N2L 3G1, Canada} 
\affiliation{Dept. Applied Math., University of Waterloo, Waterloo, ON, N2L 3G1, Canada}


\bigskip
\begin{abstract}

We show that uniformly accelerated detectors can display genuinely thermal features even if the Kubo-Martin-Schwinger (KMS) condition fails to hold. These features include satisfying thermal detailed balance and having a Planckian response identical to cases in which the KMS condition is satisfied. In this context, we discuss that satisfying the KMS condition for accelerated trajectories is just sufficient but not necessary for the Unruh effect to be present in a given quantum field theory. Furthermore, we extract the necessary and sufficient conditions for the response function of an accelerated detector to be thermal in the infinitely adiabatic limit. This analysis provides new insights about the interplay between the KMS condition and the Unruh effect, and  a solid framework in which the robustness of the Unruh effect against deformations of quantum field theories (perhaps Lorentz-violating) can be answered unambiguously.

\end{abstract}

\maketitle

\noindent
Quantum field theory (QFT) is considered to be an effective theory that is valid outside the quantum gravity domain, typically defined in terms of a length scale $\ell$ \cite{Garay1994}. This theoretical framework is therefore expected to become less precise as this scale is approached, perhaps eventually failing completely as a correct description of nature. In most cases, there is a hierarchy of scales that ensures that the predictions of QFT within its domain of validity are not contaminated by the ultraviolet physics below $\ell$. However, it is known that ultraviolet deformations of the structure of QFT can percolate into this domain of validity, spoiling the decoupling of scales \cite{Collins2004,Collins2006,Gambini2011,Polchinski2011,Belenchia2016}. When present, this phenomenon brings the possibility of testing theoretical frameworks which would be otherwise impossible to probe. Examples of this behavior that have been recently discussed in the literature include the response of particle detectors along inertial trajectories in the framework of polymer quantization \cite{Kajuri2015,Husain2015,Louko2017,Kajuri2017} and in non-local field theories \cite{Belenchia2016b}, or the transmission of information through non-local fields \cite{Belenchia2017}. Determining the deformations that lead to this percolation, and finding the predictions which are affected, is of clear importance for quantum gravity phenomenology. 

Here, we focus on a central prediction of QFT: the Unruh effect \cite{Fulling1972,Davies1974,Unruh1976}. This well-known phenomenon \cite{Crispino2007} illustrates that the concept of particle is observer-dependent in QFT, an observation that is inextricably linked to black hole evaporation \cite{Hawking1974,Hawking1974b}. Aside from its importance on theoretical grounds, there are reasonable prospects for detecting this effect in the near future (e.g., \cite{Schutzhold2008,Aspachs2010,MartinMartinez2010,MM2013}). In standard Lorentz-invariant QFT, the thermal behavior of the response function of uniformly accelerated detectors holds exactly under the Kubo-Martin-Schwinger (KMS) condition \cite{Kubo1957,Martin1959,Haag1967} that characterizes thermal states. However, and as we make explicit below, large sets of deformations of QFT (including the introduction of a cutoff on spatial momenta) lead to violations of the KMS condition. There are two possible attitudes with respect to this observation. The first one is to assume directly that these deformations erase any trace of the Unruh effect (see, for instance, \cite{Hossain2015} for a particular example in the framework of polymer quantization). The second one, put forward in this letter, is admitting that the KMS condition is unnecessarily restrictive from a physical perspective. In the presence of deformations of QFT with typical length scale $\ell$, it is reasonable to expect that small deviations from an exact thermal behavior, involving this new scale, would appear. This broader set of scenarios cannot be characterized by the KMS condition, which will be generally violated even though the response function can display thermal features. This has been noticed before in particular scenarios. For example, the thermalization of accelerated detectors to temperatures proportional to their acceleration has been described in studies involving cavities \cite{Brenna2013} where the KMS condition is not satisfied. Indeed, there are non-KMS examples where transient non-thermal behaviour is drawn out in the infinitely adiabatic limit, e.g., \cite{Louko1998}.

Following this intuition, here we determine the minimal requirements that single out the scenarios in which the violations of the KMS condition are mild enough so that the Unruh effect is preserved. We first prove that the long-time response function of a uniformly accelerated detector interacting with fields that are invariant under spacetime translations and spatial rotations (but not necessarily Lorentz boosts) is reduced, in the adiabatic limit, to a single-variable integral of the sum of the residues of the poles of the Wightman function inside a horizontal strip of the complex plane. 
This general result permits us to calculate explicitly the response function in a generality of situations, allowing us to critically revise cases of particular deformations studied previously \cite{Agullo2008,Rinaldi2008,Nicolini2009,Campo2010,Gutti2010,Agullo2010,Hossain2014,Hossain2015,Alkofer2016}, and define in general terms the conditions for the preservation of the Unruh effect. Crucially, we find that the preservation of the Unruh effect is less restrictive than the KMS condition. 

\noindent
\textsl{Deformations of the Wightman function.--}A relativistic QFT is given by a Hilbert space $\mathbb{H}$ of states and a set of unitary operators associated with the transformations in the Poincar{\'e} group. $\mathbb{H}$ has to contain a unique state $|0\rangle$ invariant under Poincar{\'e} transformations. Field operators $\phi(X)$ are operator-valued distributions acting on the space of test functions defined over $\mathbb{R}^n$.  
The two remaining conditions are energy positivity and locality (field operators commute on space-like intervals).

Wightman functions contain the full information about a QFT satisfying the axioms above \cite{Wightman1956}.  For our purposes here, it is enough to study the two-point Wightman function \mbox{$\mathscr{W}(X'',X')=\langle0|\phi(X'')\phi(X')|0\rangle$}, since the leading order detector response for any state is only function of this quantity through the so-called response function:
\begin{align}
\mathscr{F}(\Omega,\sigma) 
=\frac{1}{\sigma}\int_{-\infty}^{\infty}\text{d}\tau''&\int_{-\infty}^\infty\text{d}\tau'\,\chi(\tau''/\sigma)\chi(\tau'/\sigma)\nonumber\\
&\times W(\tau'',\tau')e^{-\ii\Omega(\tau''-\tau')}.\label{eq:basint}
\end{align}
Here, $W(\tau'',\tau')$ is the pull-back of the Wightman function $\mathscr{W}(X'',X')$ to a given trajectory in spacetime, $X(\tau)$. Eq. \eqref{eq:basint} arises naturally in the study of the excitation and decay probabilities in the Unruh-de Witt model of a detector interacting with the field $\phi(X)$ (see \cite{Fewster2016}, for instance, for a detailed description), which assumes an interaction Hamiltonian $H_{\rm I}(\tau)=\lambda\chi(\tau/\sigma)\mu(\tau)\phi(X(\tau))$, where $\mu(\tau)$ is the monopole moment operator of the detector, and $\chi(\tau/\sigma)\in C^{\infty}(\mathbb{R})$ is a square-integrable switching function that controls the duration and the form of the window of time in which the interaction between the detector and the field takes place. This kind of switching is known as adiabatic \cite{Fewster2016,Garay2016}, and it depends on a single width parameter $\sigma$ that provides a measure of the interaction time scale (so that the infinitely adiabatic limit $\sigma\rightarrow\infty$ corresponds to a detector switched on forever). The probabilities of excitation and decay are proportional to Eq. \eqref{eq:basint}, with respectively positive and negative values of $\Omega$ (the energy gap of the detector is $|\Omega|$). However, the corresponding proportionality factors are independent on the properties of the QFT, and only depend on the properties of the detector (including its monopole moment) and the coupling $\lambda\in\mathbb{R}$ between the detector and the field \cite{Birrell1982,Takagi1986}. Hence, the quotients of these probabilities, which are the quantities of interest in order to determine whether or not the response of the detector is thermal, are independent of these proportionality factors.

The Wightman function is a distribution, so that Eq. \eqref{eq:basint} would be meaningful only for suitable choices of the space of switching functions such as, for instance, functions with non-compact support that decay faster than any polynomial. For our purposes here, it will be enough to consider Gaussian switching functions normalized so that $\int_{-\infty}^\infty\text{d}y\,\chi(y)^2=1$, though other choices are possible. Hence, we can safely forget about the distributional nature of the Wightman function (and its deformations introduced below), and work with it as if it was a function, as long as we keep in mind that this quantity is always under the integral sign in Eq.  \eqref{eq:basint}.

On general grounds, the introduction of an additional length scale $\ell$ leads to deformations of the functional form of the Wightman function. These deformations encode the leading modifications arising from the particular ultraviolet completion chosen, or may just represent physical cutoffs. Let us make the following technical assumptions:
\begin{enumerate}
\item{There is an effective continuum flat description of spacetime in which the deformed Wightman function can be written as a function of the spacetime coordinates $X^\mu=(t,\bm{x})$.}
\item{The deformed Wightman function reduces to its standard Poincar{\'e} invariant form $\mathscr{W}_0(\Delta X)$ in the formal limit $\ell\rightarrow 0$.}
\item{The functional form of the deformed Wightman function may break explicitly the invariance under Lorentz boosts, while keeping spacetime translations and spatial rotations as symmetries.}
\end{enumerate}
These are fairly general assumptions. For instance, condition 1 above permits to include in our analysis discrete or quantum-mechanical features of the spacetime structure. On the other hand, we can exploit conditions 2 and 3 in order to write the deformed Wightman function as
\begin{align}
\mathscr{W}_\ell(\Delta t,\Delta \bm{x})=\mathscr{W}_0(\Delta X)[1+\mathscr{D}_\ell(\Delta t,\Delta \bm{x})].\label{eq:gendef}
\end{align}
We have made use of the fact that the transformation rules for the vacuum state and field operators imply that the Wightman function is invariant under translations. Hence, the Wightman function is a function of the differences $\Delta t=t''-t'$ and $\Delta \bm{x}=\bm{x}''-\bm{x}'$. The notation used makes explicit that invariance under boosts is not assumed.

The function $\mathscr{D}_\ell(\Delta t,\Delta \bm{x})$ has been introduced on phenomenological grounds and satisfies $\lim_{\ell\rightarrow0}\mathscr{D}_\ell(\Delta t,\Delta \bm{x})=0$. Its specific form will depend on the particular deformation that is chosen; we will discuss some examples below. But our goal here is to keep the discussion as general as possible, so that no further constraints are imposed on this quantity for the moment.

\noindent
\textsl{Infinitely adiabatic limit for uniformly accelerated observers.--}The response function $\mathscr{F}_\ell(\Omega,\sigma)$ associated with a given deformation takes the same form as  Eq. \eqref{eq:basint} but with $W(\tau'',\tau')$ replaced with $W_\ell(\tau'',\tau')$. In the following, we consider only  trajectories with constant acceleration $a$ and Gaussian switching functions. Changing the integration variables to $w=\tau''+\tau'$ and $z=\tau''-\tau'$, and performing the integration in the latter, one arrives to a more transparent relation:
\begin{align}
    &\mathscr{F}_{\ell}(\Omega,\infty)=\frac{\ii\sqrt{\pi}}{\displaystyle1-e^{4\pi\Omega/a}}
    \nonumber\\
    &\quad\times\lim_{\sigma\rightarrow\infty}\frac{1}{\sigma}\int_{-\infty}^{\infty}\text{d}w\,e^{-w^2/\sigma^2}\sum_{k\in I}\mbox{Res}[f_{\ell,\sigma}(z),z_k].\label{eq:finexp}
\end{align}
In this equation, $f_{\ell,\sigma}(z)$ is the function $f_{\ell,\sigma}(z)=e^{-z^2/\sigma^2}W_\ell(w,z)e^{-\ii\Omega z}$, with the slight abuse of notation $W_\ell(w,z)=W_\ell(\tau''(w,z),\tau'(w,z))$. On the other hand, $\{z_k\}_{k\in I}$ is the (finite) set of poles in $z$ of the Wightman function [equivalently, $f_{\ell,\sigma}(z)$] on the horizontal strip of the complex plane $S\subset\mathbb{C}$ defined by $0\leq\mbox{Im}[z]\leq 4\pi/a$.

A couple of technical remarks are needed. To write Eq. \eqref{eq:finexp}, we have exploited the periodicity properties of hyperbolic functions in order to choose an appropriate integration contour in the complex plane. Indeed, the pull-back of the Wightman function satisfies \mbox{$W_\ell(w,z+4\pi\ii/a)=W_\ell(w,z)$}. In addition, we need the following fall-off condition:
\begin{itemize}
\item[4.]{The deformed Wightman function is polynomially bounded in $|\Delta t|$ and $|\Delta \bm{x}|$ when these absolute values tend to infinity.}
\end{itemize}
Moreover, the Wightman function is generally singular on the real axis in the coincidence limit. This is typically dealt with introducing a regulator $z+\ii\epsilon$. This can be also understood as an infinitesimal displacement of one of the proper times $\tau'$ or $\tau''$ (more details in this regard are given later). This regulator is removed in the final expression for the response function after integration. We assume that all the real poles in $z$ are regularized in the same way. In deformations that are Lorentz-breaking, additional real poles in $w$ can appear. In this case, the infinitesimal displacement of the proper times $\tau'$ or $\tau''$ lead to $w\pm \ii\epsilon$. Physical results should not depend on the sign of the regulator in $w$.

\noindent
\textsl{Preservation of the Unruh effect.--}Eq. \eqref{eq:finexp} determines the infinitely adiabatic limit of the response function for all the deformations satisfying the requirements 1-4. We have used it in order to calculate the response function in several examples. The results are compiled in Table \ref{tab:1}. Most importantly, we want to highlight that this expression can be exploited in order to extract the conditions that guarantee the preservation of the Unruh effect. 

On general grounds, the Unruh effect is preserved if the response rate of the detector along uniformly accelerated trajectories has the right $\ell\rightarrow0$ limit, namely if $\lim_{\ell\rightarrow0}\mathscr{F}_{\ell}(\Omega,\infty)=\mathscr{F}(\Omega,\infty)$. This can be alternatively defined in terms of a commutative diagram involving the double integration in $w$ and $z$ and the $\ell\rightarrow0$ limit in the response function $\mathscr{F}_\ell(\Omega,\infty)$. Note that, in the infinitely adiabatic limit, the only possible dimensionless combinations of the physical quantities involved are $\ell a$ and $\ell\Omega$. Hence, if this condition is satisfied, the $\ell=0$ expressions for the response functions are recovered up to small corrections when $\ell a\ll1$ and $\ell\Omega\ll1$. In other words, appreciable deviations from the Unruh effect would only exist for accelerations or frequencies that are of the same scale as the inverse of the parameter of the deformation $\ell$. This realizes the decoupling of scales that we alluded to in the introduction. 

It is worth mentioning that the study of the finite-time response, and not only its adiabatic limit, is a source of rich phenomenology (see among others \cite{Langlois2005,Rovelli2011,Belenchia2016b,Belenchia2017}). This finite-time response would be sensitive to the details of the switching, including additional physical scales that appear in non-adiabatic switching functions. These additional scales may form new dimensionless combinations with high-energy scales that may not be necessarily small. This is just a reminder of the following fact: that two different deformations preserve the Unruh effect does not necessarily imply that all other possible observables will agree, and therefore that these are observationally indistinguishable.

We can now use Eq. \eqref{eq:finexp} in order to identify the necessary and sufficient conditions that guarantee the preservation of the Unruh effect. Let us define the set of poles $\{\bar{z}_l\}_{l\in J}$ that are obtained as a continuous deformation of the original set of poles $\{z^0_m\}_{m\in K}$ originally in $S\subset\mathbb{C}$, with the possible addition or splitting of poles. Let us start with a necessary condition:

\begin{itemize}
\item[A.]{\emph{Local uniform convergence:} The integral of the response function along each of the contours $\gamma_i$ containing all the deformed poles that stem from each of the poles $z^0_i$ of the undeformed Wightman function, but not from other poles, has the right $\ell\rightarrow0$ limit.}
\end{itemize}
If this condition holds, it is possible to identify two further necessary conditions: 
\begin{itemize}
\item[B.]{The sum of the residues of the relevant poles of the Wightman function in $z$ must be integrable with respect to $w$ in the $\sigma\rightarrow\infty$ limit.}
\item[C.]{All these poles must remain in the horizontal strip $S\subset\mathbb{C}$, namely $\{z_k\}_{k\in I}=\{\bar{z}_l\}_{l\in J}$.}
\end{itemize}
These three necessary conditions (A,B,C) are in fact sufficient, when holding simultaneously, in order to preserve the Unruh effect.

Let us sketch the proof of this statement. Condition B implies that the right-hand side of Eq. \eqref{eq:finexp} is finite in the $\sigma\rightarrow\infty$ limit. On the other hand, condition C implies that all the deformed poles that stem from undeformed poles $z^0_m$ inside the horizontal strip $S\subset\mathbb{C}$ remain in $S$. Therefore, the corresponding residues are all taken into account in the right-hand side of Eq. \eqref{eq:finexp}. Finally, condition A ensures that the sum of these residues has the right $\ell\rightarrow0$ limit.

\noindent
\textsl{Regarding the Kubo-Martin-Schwinger condition.--} As mentioned previously, in Lorentz-invariant QFT the thermal behavior of the response function holds exactly under the KMS condition. In fact, it can be seen \cite{Fewster2016} that this condition is sufficient for (thermal) detailed balance to be satisfied,
\begin{equation}
\mathscr{F}(-\Omega,\infty)=e^{2\pi\Omega/a}\mathscr{F}(\Omega,\infty),\label{eq:detbal}
\end{equation}
which is the smoking gun of thermal behavior. However, we have devoted this letter to the determination of the minimal requirements that single out the scenarios in which this violation is mild enough so that detailed balance is still satisfied, possibly up to small corrections. For completeness, in this section we discuss the interplay between these different conditions, and illustrate the general discussion with examples (see Table \ref{tab:1}).

First of all, let us clarify the operational definition of the KMS condition that we will be using \cite{Strocchi2008,Fewster2016}. In fact, the KMS condition is, more strictly, a series of conditions. The first one can be defined in abstract terms as the following property, to be satisfied by any pair of operators $A=A(t=t_0)$ and $B=B(t=t_0)$ (for some arbitrary value of $t_0$) evolved in some time parameter $t$ in the Heisenberg picture: there exists some $\beta\in\mathbb{R}$ such that
\begin{equation}
\langle A(t+\ii\beta-\ii\epsilon)B\rangle=\langle BA(t-\ii\epsilon)\rangle.\label{eq:kmspdef1}
\end{equation}
We have introduced a suitable regularization (the $i\epsilon$ terms) that is needed in order to formally manipulate distributions as functions. Translated in terms of the Wightman function $W_\ell(\tau'',\tau')=\langle0|\phi(X(\tau''))\phi(X(\tau'))|0\rangle$, the equation above reads
\begin{equation}
W_\ell(\tau'+\ii\beta-\ii\epsilon,\tau'')=W_\ell(\tau'',\tau'-\ii\epsilon).\label{eq:kmsdef1}
\end{equation}
This explains why the KMS condition is sometimes defined in the literature (e.g., \cite{Rovelli2014,Hossain2015,Alkofer2016}) just as the symmetry of the pull-back of the Wightman function under the transformation $\tau'\rightarrow \tau''+\ii\epsilon$ and $\tau''\rightarrow \tau'+\ii\beta-\ii\epsilon$, with $\beta=2\pi/a$. It can be checked explicitly that this symmetry is equivalent to the imaginary periodicity $z\rightarrow z+\ii\beta$.

When Eq. \eqref{eq:kmspdef1} is satisfied by all the possible operators $A$ and $B$, including the field operators $\phi(X)$ but also the identity operator $\mathbbm{1}$, it follows that the state $|0\rangle$ must be invariant under time translations. As a consequence, the pull-back of the Wightman function must be stationary, namely invariant under translations in $\tau$ or, equivalently, a function of $z$ only. Note that Eq. \eqref{eq:kmsdef1} being satisfied does not therefore imply by itself stationarity. This is especially important for the present discussion, as it is not difficult to show that the pull-back to uniformly accelerated trajectories of any explicitly Lorentz-violating Wightman function cannot be stationary, and therefore it must necessarily violate the KMS condition (the stationarity condition can still be satisfied on inertial trajectories \cite{Husain2015,Louko2017}).

However, the KMS condition involves additional restrictions \cite{Strocchi2008,Fewster2016}, which are not always emphasized in the literature. The pull-back of the Wightman function must be holomorphic in $z$ in a horizontal strip of the lower complex semi-plane with a width $2\pi /a$ in the $\epsilon\rightarrow0$ limit and the real axis being one of its boundaries (under the condition of imaginary periodicity above, the position of this horizontal strip can be shifted by an arbitrary multiple of $2\pi \ii/a$). Therefore, it is not only necessary to show that Eq. \eqref{eq:kmsdef1} holds, but also the absence of poles inside this horizontal strip must be shown in order to claim that the KMS condition holds (equivalently, any poles must be located in the boundary of the strip in the $\epsilon\rightarrow0$ limit). Lastly, there is another further condition, which is similar but more restrictive than our condition 4: the pull-back of the Wightman function inside the complex strip must be polynomially bounded. 

These are all the ingredients that are needed in order to compare the KMS condition with the definition for the preservation of the Unruh effect given in this letter. This is summarized in the Table \ref{tab:1} below. There are several aspects which are worth stressing. The first one is that the KMS condition is not necessary for the preservation of the Unruh effect. That is, as anticipated in the introduction of this letter, it is a sufficient but not necessary condition. In particular, violating the imaginary periodicity is not sufficient in order to claim that the Unruh effect is not present. The second observation is about the interplay between the imaginary periodicity of the Wightman function and stationarity. It may seem surprising that there are deformations that satisfy the former but not the latter. However, as discussed above, stationarity follows from the stronger condition of imaginary periodicity for arbitrary pairs of operators (including the identity $\mathbbm{1}$). Moreover, it is not difficult to see that this kind of behavior is quite general, as one can show that
\begin{align}
&W_\ell(\tau'+\ii\beta-\ii\epsilon,\tau'')\nonumber\\
&=\mathscr{W}_\ell(\Delta t(\tau'+\ii\beta-\ii\epsilon,\tau''),\Delta \bm{x}(\tau'+\ii\beta-\ii\epsilon,\tau''))\nonumber\\
&=\mathscr{W}_\ell(-\Delta t(\tau'',\tau'-\ii\epsilon),-\Delta \bm{x}(\tau'',\tau'-\ii\epsilon)).
\end{align}
The first identity just makes explicit that the pull-back of the Wightman function depends on $\tau'$ and $\tau''$ implicitly through the time and space intervals, while the second identity exploits the periodicity properties of hyperbolic functions. It follows that any Wightman function that is quadratic in the time and space intervals satisfies the imaginary periodicity condition.

\begin{widetext}
\begin{center}
\begin{table}[H]
\begin{center}
\begin{tabular}{|c|c|c|c|c|c|}
\hline
 $\mathscr{D}_\ell(\Delta t,\Delta \bm{x})$ & \multicolumn{4}{c}{KMS}\vline & Preservation\\
\hline
& Imaginary periodicity &Stationarity & Holomorphicity & Polynomial & \\
\hline
$\ell^2/(\Delta X^2+\ell^2)$ \cite{Agullo2008,Campo2010} & \cmark & \cmark & \cmark & \cmark & \cmark\\ 
 \hline
$-\beta\ell^2/(\Delta X^2+\ell^2)$ & \cmark & \cmark  & \cmark & \cmark & \cmark \\
 \hline  
$-\ell^2/(\Delta X^2-\ell^2)$ \cite{Campo2010} & \cmark & \cmark & \xmark & \cmark & \xmark \\
\hline
$-e^{-\Delta X^2/\ell^2}$ \cite{Nicolini2009,Rinaldi2010} & \cmark & \cmark & \cmark & \xmark & \textbf{?} \\
\hline
$\ell^2/\Delta t^2$ & \cmark & \xmark & \cmark & \cmark & \cmark \\
\hline
 $\ell/(\Delta t-\ell)$ & \xmark & \xmark & \cmark & \cmark & \xmark \\
\hline
$\ii \ell\Delta t/\Delta X^2$ & \xmark & \xmark & \cmark & \xmark & \xmark \\
\hline   
$\ii\ell/\Delta t$ & \xmark & \xmark & \cmark & \cmark & \cmark \\
\hline   
\end{tabular}
\caption{Comparison of the KMS condition and the sufficient conditions for the preservation of the Unruh effect in the adiabatic limit. Note that ``Imaginary periodicity'' refers to the property of the Wightman function alone, namely Eq. \eqref{eq:kmsdef1}, and not the more general Eq. \eqref{eq:kmspdef1}. The latter can be satisfied only if stationarity holds.}
\label{tab:1}
\end{center}
\end{table}
\end{center}
\end{widetext}

\noindent
\textsl{Conclusions.--}We have analyzed the interplay between the conditions that guarantee the appearance of the Unruh effect and the KMS condition. We have shown that the latter is more restrictive, as it is a sufficient, but not necessary, condition to ensure that the response of a uniformly accelerated detector displays a thermal behavior. The importance of this observation is better understood if we take into account that the KMS condition was the focus of the analysis of the Unruh effect in previous works. In order to illustrate that focusing on the KMS condition is not an adequate approach, we have provided explicit examples in which the KMS condition is violated in different ways, while the response of a detector in the adiabatic limit still displays a thermal behavior. Thus, in these scenarios, no adiabatic thermalization experiment will find any contradiction with the Unruh effect despite the KMS violation. Our analysis settles the practical issue of determining whether or not a particular deformation of a QFT preserves the Unruh effect, providing the necessary tools to answer this question in a wide range of scenarios.

\acknowledgments

The authors thank Jorma Louko and Stefano Liberati for helpful discussions. This work has been supported in part by the MINECO (Spain) projects FIS2014-54800-C2-2-P and  FIS2017-86497-C2-2-P (with FEDER contribution). The work of E.M.-M. is supported by the Natural Sciences and Engineering Research Council of Canada through the Discovery program. E.M.-M. also gratefully acknowledges the funding of his Ontario Early Research Award.

\bibliography{refs}	

\end{document}